\documentclass[twocolumn,aps,pra,showpacs]{revtex4}
\usepackage[T1]{fontenc}
\usepackage[latin9]{inputenc}
\usepackage{bm}
\usepackage{amsmath}
\usepackage{amssymb}
\usepackage{graphicx}
\usepackage{esint}

\makeatletter

\newcommand*\LyXThinSpace{\,\hspace{0pt}}

\@ifundefined{textcolor}{}
{%
 \definecolor{BLACK}{gray}{0}
 \definecolor{WHITE}{gray}{1}
 \definecolor{RED}{rgb}{1,0,0}
 \definecolor{GREEN}{rgb}{0,1,0}
 \definecolor{BLUE}{rgb}{0,0,1}
 \definecolor{CYAN}{cmyk}{1,0,0,0}
 \definecolor{MAGENTA}{cmyk}{0,1,0,0}
 \definecolor{YELLOW}{cmyk}{0,0,1,0}
}

\makeatother

\begin{document}

\title{Anderson localization of Cooper pairs and Majorana fermions in an
ultracold atomic Fermi gas with synthetic spin-orbit coupling}

\author{Ye Cao$^{1}$, Gao Xianlong$^{2}$, Xia-Ji Liu$^{1}$, and Hui Hu$^{1}$}

\affiliation{$^{1}$Centre for Quantum and Optical Science, Swinburne University
of Technology, Melbourne 3122, Australia}

\affiliation{$^{2}$Department of Physics, Zhejiang Normal University, Jinhua
321004, China}

\date{\today}
\begin{abstract}
We theoretically investigate two-particle and many-particle Anderson
localizations of a spin-orbit coupled ultracold atomic Fermi gas trapped
in a quasi-periodic potential and subjected to an out-of-plane Zeeman
field. We solve exactly the two-particle problem in a finite length
system by exact diagonalization and solve approximately the many-particle
problem within the mean-field Bogoliubov-de Gennes approach. At a
small Zeeman field, the localization properties of the system are
similar to that of a Fermi gas with conventional $s$-wave interactions.
As the disorder strength increases, the two-particle binding energy
increases and the fermionic superfluidity of many particles disappears
above a threshold. At a large Zeeman field, where the interatomic
interaction behaves effectively like a $p$-wave interaction, the
binding energy decreases with increasing disorder strength and the
resulting topological superfluidity shows a much more robust stability
against disorder than the conventional $s$-wave superfluidity. We
also analyze the localization properties of the emergent Majorana
fermions in the topological phase. Our results could be experimentally
examined in future cold-atom experiments, where the spin-orbit coupling
can be induced artificially by using two Raman lasers, and the quasi-periodic
potential can be created by using bichromatic optical lattices. 
\end{abstract}


\maketitle
Ultracold atomic Fermi gases provide one of the most versatile platforms
for realizing exotic many-body quantum states of matter, due to their
unprecedented tunability and controllability \cite{Bloch2008,Giorgini2008}.
Strongly interacting Fermi gases are easily achievable through the
use of Feshbach resonances \cite{Chin2010}, which enable the realization
of the crossover from Bose-Einstein condensates (BEC) to Bardeen-Cooper-Schrieffer
(BCS) superfluids \cite{Giorgini2008}. Low-dimensional Fermi gases
are accessible by imposing optical lattice potentials \cite{Liao2010,Martiyanov2010,Pagano2014},
potentially allowing the observation of the elusive antiferromagnetic
Néel order in the fermionic Hubbard model \cite{Hart2015}. Two more
examples are the recently synthesized spin-orbit coupled Fermi gases
\cite{Wang2012,Cheuk2012,Fu2013} and quasi-disordered Fermi gases
in bichromatic optical lattices \cite{Schreiber2015,Bordia2015},
which open the ways to realize topological superfluids \cite{Liu2012a,Liu2012b,Wei2012,Zhang2014}
and many-body localizations \cite{Schreiber2015,Bordia2015,Basko2006},
respectively.

In this Rapid Communication, motivated by these rapid experimental
progress, we propose to investigate Anderson localization of two and
many interacting fermions in one-dimensional (1D) quasi-disordered
lattices, in the presence of a synthetic spin-orbit coupling and an
out-of-plane Zeeman field. By increasing the strength of the Zeeman
field above a threshold, the underlying character of the effective
interatomic interaction changes from $s$-wave to $p$-wave \cite{Zhang2008},
and in the absence of disorder the many-fermion system is known to
experience a topological phase transition \cite{Liu2012a,Liu2012b,Wei2012}.
The main purpose of this work is to determine a rich phase diagram
due to the interplay between disorder and $s$-wave or $p$-wave superfluidity.

\begin{figure}
\begin{centering}
\includegraphics[width=0.45\textwidth]{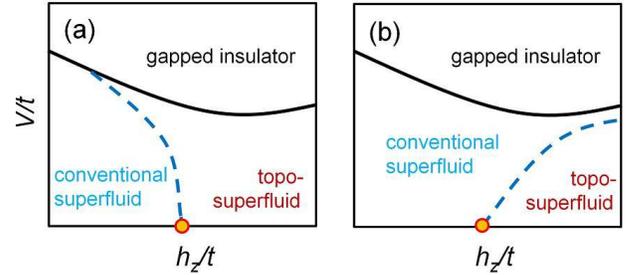} 
\par\end{centering}

\protect\caption{(color online) Two possible phase diagrams for a spin-orbit coupled
Fermi gas in a disordered potential. In the first-type diagram (a),
as the disorder strength increases, a conventional $s$-wave superfluid
turns into a topological superfluid before losing its superfluidity.
The transition from a topological superfluid to a gapped insulator
is direct. In contrast, in the second-type diagram (b), with increasing
disorder strength a topological superfluid first becomes a conventional
superfluid and then translates into a gapped insulator.}

\label{fig1}
\end{figure}

Under sufficiently strong disorder potential, a dirty $s$-wave superconductors
such as a superconducting NbN thin film undergoes a quantum phase
transition towards a gapped insulator \cite{Mondal2011}. In condensed
matter physics, such a superconductor-insulator transition has attracted
a lot of interests over the past two decades \cite{Ma1985,Ghosal2001,Feigelman2007,Dubi2007,Bouadim2011,Seibold2012}.
Yet, a complete understanding of the transition remains elusive. Also,
the behavior of a dirty $p$-wave superconductor and its phase transition
towards a gapped insulator is less known \cite{DeGottardi2013,Cai2012,Wang2015}.
Our proposed system of disordered spin-orbit coupled atomic Fermi
gases offers an unique opportunity to better understand these quantum
phase transitions. In particular, it is of great interest to determine
how does disorder affect the topological phase transition and how
is the fate of Majorana fermion edge modes - as the key feature of
a topological superfluid - against disorder. Naïvely, we may anticipate
two different phase diagrams, as schematically illustrated in Fig.
\ref{fig1}. As the disorder strength increases, the topological phase
transition line (the dashed curve) could bend to either the left or
right hand side of the circle that indicates the initial transition
point in the clean limit. In other words, with increasing disorder,
an $s$-wave superfluid may first become a topological superfluid
before it finally turns into a gapped insulator (i.e., the scenario
a), or vice versa (b).

The proposed 1D spin-orbit coupled Fermi system in quasi-disordered
lattices can be easily realized in cold-atom laboratory, where the
synthetic spin-orbit coupling and disorder potential can be created
by using Raman laser beams and bichromatic optical lattices \cite{Zhou2013},
respectively. By using the self-consistent Bogoliubov-de Gennes mean-field
theory, we calculate the superfluid density (under periodic boundary
condition) and the Majorana fermion edge modes (with open boundary
condition) of the system at zero temperature. A vanishingly small
superfluid density indicates the transition to a gapped insulator,
while the existence of Majorana fermions signals the topological phase
transition. This allows us to determine the whole phase diagram (see
Fig. \ref{fig4}), the main result of this work. To understand the
phase diagram, we exactly solve the two-fermion problem and address
the Anderson localization of a Cooper pair in its ground state \cite{Dufour2012}.
To utilize and better understand the robustness of Majorana fermions
with respect to disorder, we also consider the soliton-induced Majorana
fermions \cite{Xu2014,Liu2015} and investigate the evolution of their
energies and wave-functions as a function of the disorder strength.
Our results on the Anderson localization of Majorana fermions may
be useful for manipulating these exotic non-abelian quasi-particles
in realistic noisy environment, for the purpose of performing fault-tolerant
quantum information processing \cite{Nayak2008}.

\textit{Model Hamiltonian}. --- We start by considering a 1D disordered
spin-orbit coupled Fermi gas in a lattice with $L$ sites. The system
can be described by the fermionic model Hamiltonian,

\begin{eqnarray}
\mathcal{H} & = & -t\sum_{i=1}^{L}\sum_{\sigma}\left(\bm{c}_{i+1,\sigma}^{\dagger}\bm{c}_{i,\sigma}+\textrm{H.c.}\right)+\sum_{i,\sigma}V_{i}\bm{n}_{i\sigma}\nonumber \\
 &  & +\mathcal{H}_{R}+\mathcal{H}_{Z}-U\sum_{i}\bm{n}_{i\uparrow}\bm{n}_{i\downarrow},\label{eq:ModelHami}
\end{eqnarray}
where $\bm{c}_{i\sigma}$ is the annihilation operator with spin $\sigma\in\{\uparrow,\downarrow\}$
at site $i$, $\bm{n}_{i\sigma}=\bm{c}_{i\sigma}^{\dagger}\bm{c}_{i\sigma}$
is the local number operator, $t$ is the tunneling strength between
neighboring lattice sites, $V_{i}=V\cos(2\pi\beta i+\phi)$ with an
irrational number $\beta$ and phase offset $\phi$ describes the
quasi-random disorder potential induced by bichromatic (i.e., additional
incommensurate) lattices \cite{Schreiber2015,Bordia2015}, and $U>0$
in the last term represents the on-site attractive interaction. For
definiteness, we shall take $\beta=(\sqrt{5}-1)/2$ and $\phi=0$.
We assume a periodic boundary condition such that $\bm{c}_{L+1,\sigma}=\bm{c}_{1\sigma}$,
unless specified otherwise. Finally, the spin-orbit term with Rashba-type
coupling $\mathcal{H}_{R}$ and the Zeeman energy term $\mathcal{H}_{Z}$
are given by, 
\begin{eqnarray}
\mathcal{H}_{R} & = & \frac{\lambda}{2}\sum_{i=1}^{L}\left(\bm{c}_{i+1,\downarrow}^{\dagger}\bm{c}_{i\uparrow}-\bm{c}_{i+1,\uparrow}^{\dagger}\bm{c}_{i\downarrow}+\textrm{H.c.}\right),\\
\mathcal{H}_{Z} & = & h_{z}\sum_{i=1}^{L}\left(\bm{c}_{i\uparrow}^{\dagger}\bm{c}_{i\uparrow}-\bm{c}_{i\downarrow}^{\dagger}\bm{c}_{i\downarrow}\right),
\end{eqnarray}
respectively. Here, $\lambda$ is the spin-orbit coupling strength
and $h_{z}$ is the Zeeman field. 

We note that, in the absence of $\mathcal{H}_{R}$ and $\mathcal{H}_{Z}$,
the model Hamiltonian Eq. (\ref{eq:ModelHami}) - known as Aubry-André-Harper
model \cite{Harper1955,Aubry1980} - has been experimentally explored
\cite{Schreiber2015}. In the non-interacting limit ($U=0$), all
single-particle states become localized at the same critical disorder
strength $V=2t$ \cite{Aubry1980}. For an initial charge density-wave
state, its many-body localization at arbitrary $U$ has been demonstrated
\cite{Schreiber2015}. The possibility of having the spin-orbit and
Zeeman terms in the model Hamiltonian was recently proposed and derived
by Zhou and co-workers \cite{Zhou2013}. The Anderson localization
of a single atom in the non-interacting limit ($U$=0) was studied.
It was found that the spin-orbit coupling can lead to a non-pure spectrum
(i.e., coexistence of extended and localized states) and the appearance
of mobility edges. Here, we are interested in the localization of
the many-body ground state and the interplay between disorder and
conventional/unconventional superfluidity.

\textit{Localization of a single Cooper pair}. --- Before discussing
the localization of the coherent (superfluid) state of many Cooper
pairs, it is instructive to understand the localization of a single
Cooper pair \cite{Dufour2012}. For this purpose, we numerically exactly
solve the two-fermion problem:
\begin{equation}
\left(\mathcal{H}-E\right)\sum_{i,j}\left[\psi_{ij}^{\uparrow\uparrow}\bm{c}_{i\uparrow}^{\dagger}\bm{c}_{j\uparrow}^{\dagger}+\psi_{ij}^{\uparrow\downarrow}\bm{c}_{i\uparrow}^{\dagger}\bm{c}_{j\downarrow}^{\dagger}+\psi_{ij}^{\downarrow\downarrow}\bm{c}_{i\downarrow}^{\dagger}\bm{c}_{j\downarrow}^{\dagger}\right]\left|\textrm{0}\right\rangle =0,
\end{equation}
where $\psi_{ij}^{\sigma\sigma'}$ is the two-particle wave-function
and $E$ is the energy. It is understood that for the same spin $\sigma=\sigma'$
at the same site $i=j$, $\psi_{ij}^{\sigma\sigma'}=0$ because of
Pauli exclusion principle. For numerical stability, we approximate
the irrational number $\beta$ as the limit of a continued fraction,
$\beta\simeq F_{l-1}/F_{l}$, where $F_{l}$ are Fibonacci numbers
and $l$ is a sufficiently large integer. We minimize the finite-size
effect by taking the length of the lattice $L=F_{l}$ \cite{Dufour2012}.
To characterize the Anderson localization of the pair, we use the
inverse participation ratio (IPR) \cite{Evers2008},
\begin{equation}
\alpha_{\textrm{IPR}}=\sum_{i,j}\left(\left|\psi_{ij}^{\uparrow\uparrow}\right|^{4}+\left|\psi_{ij}^{\uparrow\downarrow}\right|^{4}+\left|\psi_{ij}^{\downarrow\downarrow}\right|^{4}\right).
\end{equation}
For an extended state, $\alpha_{\textrm{IPR}}\propto1/L^{2}$ decreases
to zero in the thermodynamic limit. While for a localized state, it
saturates to a finite value.

\begin{figure}
\begin{centering}
\includegraphics[width=0.35\textwidth]{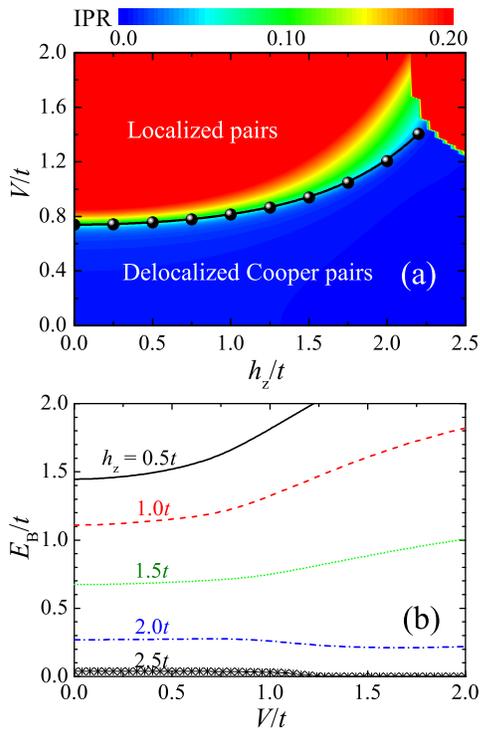} 
\par\end{centering}

\protect\caption{(color online) (a) Anderson localization of Cooper pairs, indicated
by the inverse participation ratio (IPR) of the ground state two-particle
wave-function at $L=F_{l=10}=89$. The black circles with line shows
the phase boundary in the rational limit $L\rightarrow\infty$. (b)
Binding energy of the ground state Cooper pair as a function of the
disorder strength at several Zeeman fields. Here, we set $U=4.5t$
and $\lambda=2t$.}

\label{fig2}
\end{figure}

We have calculated $\alpha_{\textrm{IPR}}$ for the ground state of
the pair at a moderate interaction $U=4.5t$ and spin-orbit coupling
$\lambda=2t$, as reported in the phase diagram Fig. \ref{fig2}a.
As the disorder strength $V$ increase, there is a sharp increase
in the IPR. We then identify the critical strength $V_{c,2p}$ as
the inflection point of the calculated curve $\alpha_{\textrm{IPR}}(V)$.
By repeating calculations at different lengths of lattices (up to
$L=F_{12}=233$), we eliminate the finite-size dependence and determine
the phase boundary $V_{c,2p}$ in the thermodynamic limit, as shown
by solid circles in the figure. It is interesting that with increasing
Zeeman field, the threshold $V_{c,2p}$ increases. This might be understood
from the fact that a finite Zeeman field plays a role of pair breaker
and the resulting weakly bound pair is easier to move in disordered
potential than a tightly bound pair \cite{Dufour2012}. In Fig. \ref{fig2}b,
we present the binding energy $E_{B}=2\epsilon_{1}-E$ of the pair
as a function of the disorder strength at different Zeeman fields.
Here $\epsilon_{1}$ is the ground state energy for a single fermion.
Indeed, with increasing Zeeman field, the binding energy decreases
quickly. As a result, at sufficiently large Zeeman field, a Cooper
pair may break into two fermions. This pair breaking effect is mostly
evident at $h_{z}=2.5t$, where the pair creases to exist at $V\gtrsim1.2t$.
In this part of parameter space (i.e., the top-right part of Fig.
\ref{fig2}a), actually we observe the Anderson localization of a
single fermion instead of a Cooper pair \cite{Zhou2013}. 

It is also worth noting that the binding energy shows very different
dependence on the disorder strength at small and large Zeeman fields.
While the disorder enhances the binding energy at low Zeeman field,
it breaks pair at high field. These two distinct behaviors may be
attributed to the different effective interatomic interactions. For
a spin-orbit coupled Fermi gas, with increasing Zeeman field, the
underlying interaction between atoms changes from $s$-wave like to
$p$-wave like with increasing the Zeeman field above a threshold
\cite{Zhang2008}. For the parameter used in Fig. \ref{fig2}b, this
transition occurs at about $h_{z}\sim1.8t$.

\textit{Mean-field phase diagram}. --- We now turn to consider the
ground state of many fermions by using mean-field Bogoliubov-de Gennes
theory, which is known to capture the qualitative physics in low dimensions
\cite{Ghosal2001}. To this aim, we decouple the interaction term
$-U\sum_{i}\bm{n}_{i\uparrow}\bm{n}_{i\downarrow}$ into the pairing
term and the Hartree-Fock term, $\mathcal{H}_{\Delta}=\sum_{i}(\Delta_{i}\bm{c}_{i\uparrow}^{\dagger}\bm{c}_{i\downarrow}^{\dagger}+\Delta_{i}^{*}\bm{c}_{i\downarrow}\bm{c}_{i\uparrow}+|\Delta_{i}|^{2}/U)$
and $\mathcal{H}_{\textrm{HF}}=\sum_{i}(-U\langle\bm{n}_{i\uparrow}\rangle\bm{n}_{i\downarrow}-U\langle\bm{n}_{i\downarrow}\rangle\bm{n}_{i\uparrow}+U\langle\bm{n}_{i\uparrow}\rangle\langle\bm{n}_{i\downarrow}\rangle)$,
which yields an effective quadratic Hamiltonian,

\begin{eqnarray}
\mathcal{H}_{\textrm{eff}} & = & -t\sum_{i\sigma}\left(\bm{c}_{i+1,\sigma}^{\dagger}\bm{c}_{i,\sigma}+\textrm{H.c.}\right)+\sum_{i,\sigma}\left(V_{i}-\tilde{\mu}_{i\bar{\sigma}}\right)\bm{n}_{i\sigma}\nonumber \\
 &  & +\mathcal{H}_{R}+\mathcal{H}_{Z}+\sum_{i}\left(\Delta_{i}\bm{c}_{i\uparrow}^{\dagger}\bm{c}_{i\downarrow}^{\dagger}+\textrm{H.c.}\right)+E_{0},\label{eq:mfHami}
\end{eqnarray}
where $E_{0}=\sum_{i}(|\Delta_{i}|^{2}/U+U\langle\bm{n}_{i\uparrow}\rangle\langle\bm{n}_{i\downarrow}\rangle)$
is a constant and $\tilde{\mu}_{i\sigma}=\mu+U\langle\bm{n}_{i\sigma}\rangle$
is the local chemical potential that incorporates the site-dependent
Hartree shift. We diagonalize the effective Hamiltonian by using the
standard Bogoliubov transformation, which leads to,
\begin{equation}
\left[\begin{array}{cccc}
\hat{K}_{+} & \hat{\Lambda}^{\dagger} & 0 & \hat{\Delta}\\
\Lambda & \hat{K}_{-} & -\hat{\Delta} & 0\\
0 & -\hat{\Delta}^{*} & -\hat{K}_{+}^{*} & -\hat{\Lambda}^{\dagger*}\\
\hat{\Delta}^{*} & 0 & -\Lambda^{*} & -\hat{K}_{-}^{*}
\end{array}\right]\left[\begin{array}{c}
u_{i\uparrow}^{(\eta)}\\
u_{i\downarrow}^{(\eta)}\\
v_{i\uparrow}^{(\eta)}\\
v_{i\downarrow}^{(\eta)}
\end{array}\right]=E_{\eta}\left[\begin{array}{c}
u_{i\uparrow}^{(\eta)}\\
u_{i\downarrow}^{(\eta)}\\
v_{i\uparrow}^{(\eta)}\\
v_{i\downarrow}^{(\eta)}
\end{array}\right],\label{eq:BdG}
\end{equation}
where $u_{i\sigma}^{(\eta)}$ and $v_{i\sigma}^{(\eta)}$ are the
Bogoliubov quasiparticle wavefunctions and $E_{\eta}$ is the associated
quasiparticle energy. The operators $\hat{K}_{\pm}=\hat{K}\pm h_{z}$,
$\Lambda$, and $\hat{\Delta}$ are respectively given by,
\begin{gather}
\hat{K}u_{i\sigma}^{(\eta)}=-t\left[e^{-i\theta}u_{i+1,\sigma}^{(\eta)}+e^{i\theta}u_{i-1,\sigma}^{(\eta)}\right]+\left(V_{i}-\tilde{\mu}_{i\bar{\sigma}}\right)u_{i\sigma}^{(\eta)},\nonumber \\
\hat{\Lambda}u_{i\uparrow}^{(\eta)}=+(\lambda/2)\left[e^{-i\theta}u_{i+1,\downarrow}^{(\eta)}-e^{i\theta}u_{i-1,\downarrow}^{(\eta)}\right],\nonumber \\
\hat{\Lambda}^{\dagger}u_{i\downarrow}^{(\eta)}=-(\lambda/2)\left[e^{-i\theta}u_{i+1,\uparrow}^{(\eta)}-e^{i\theta}u_{i-1,\uparrow}^{(\eta)}\right],\nonumber \\
\hat{\Delta}u_{i\sigma}^{(\eta)}=\Delta_{i}u_{i\sigma}^{(\eta)},
\end{gather}
and similarly for $v_{i\sigma}^{(\eta)}$. The local pairing gap and
local density are determined by the self-consistency conditions, 
\begin{gather}
\Delta_{i}=-U\sum_{\eta}\left[u_{i\uparrow}^{(\eta)}v_{i\downarrow}^{(\eta)*}f(E_{\eta})+u_{i\downarrow}^{(\eta)}v_{i\uparrow}^{(\eta)*}f(-E_{\eta})\right],\nonumber \\
\langle\bm{n}_{i\sigma}\rangle=\sum_{\eta}\sum_{\sigma}\left[\left|u_{i\sigma}^{(\eta)}\right|^{2}f\left(E_{\eta}\right)+\left|u_{i\sigma}^{(\eta)}\right|^{2}f(-E_{\eta})\right],
\end{gather}
where $f(E_{\eta})=1/(e^{E_{\eta}/k_{B}T}+1)$ is the Fermi-Dirac
distribution function. Throughout the paper, we focus on the zero-temperature
case where $f(E_{\eta})$ becomes a step function.

\begin{figure}
\begin{centering}
\includegraphics[width=0.45\textwidth]{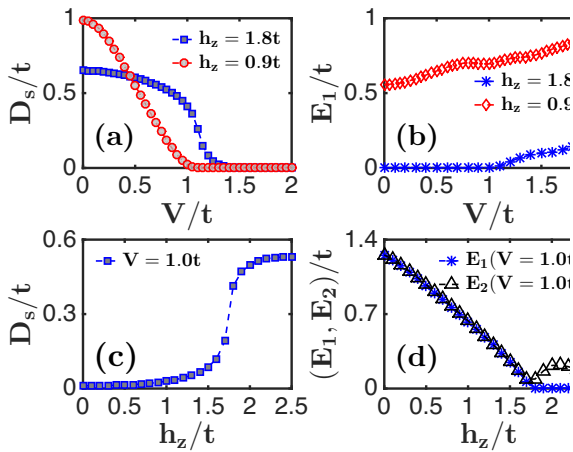} 
\par\end{centering}

\protect\caption{(color online) Left panel: Evolution of the superfluid stiffness as
a function of the disorder strength (a) or Zeeman field strength (c).
Right panel: The lowest or the second lowest quasiparticle excitation
energy as a function of the disorder strength (b) or Zeeman field
strength (d), under an open boundary condition. Here, we take $L=89$
and set $U=4.5t$ and $\lambda=2t$. Moreover, we use an average (quarter)
filling factor $n=N/L=44/89$. }

\label{fig3}
\end{figure}

We note that in defining the operators $\hat{K}$ and $\hat{\Lambda}$,
a phase twist $\theta$ is explicitly added, which corresponds to
introduce a constant vector potential $\theta_{i}=\theta$ across
the system \cite{footnote1}. This implementation is convenient for
the calculation of the superfluid stiffness $D_{s}$. Indeed, with
the vector potential, $D_{s}$ is given by the change in the free
energy $F=\left\langle \mathcal{H}_{e\textrm{ff}}\right\rangle +\mu N$
\cite{Seibold2012}:
\begin{equation}
D_{s}=\frac{1}{L}\frac{\partial^{2}F\left(\theta\right)}{\partial\theta^{2}}.
\end{equation}
It is necessary to use such a formal expression in the presence of
disorder, since the decoupling between longitudinal and transverse
electromagnetic responses does not hold and the standard approach
of calculating $D_{s}$ through the BCS response function fails \cite{Seibold2012}.

For the localization of many fermions in their ground state, we use
a vanishingly small superfluid stiffness to characterize a gapped
insulator phase. Alternatively, we may also determine the onset of
the localization phase through the calculation of the IPR of Bogoliubov
quasiparticle wavefunctions, as in the case of two fermions. However,
the use of exponentially small superfluid stiffness turns out to be
a more physical criterion. On the other hand, as we mentioned earlier,
the system features a topological superfluid at relatively large Zeeman
fields. While the existence of a non-trivial topology of the superfluid
can be revealed by some topological invariants \cite{Wei2012}, numerically
it is actually more convenient to be identified from the appearance
of two Majorana fermion modes at the edges, if one imposes an open
boundary condition.

In the left panel of Fig. \ref{fig3}, we report the superfluid stiffness
as a function of the disorder strength and of the Zeeman field. The
corresponding results for the energy of low-lying quasiparticle modes
are shown in the right panel of the figure. At a given Zeeman field
(Fig. \ref{fig3}a), the superfluid stiffness vanishes exponentially
above a critical value $V_{c}$, signifying the transition towards
a gapped insulator. The energy of low-lying modes exhibits different
dependence on disorder strength at low and high Zeeman fields (Fig.
\ref{fig3}b). At a small Zeeman field $h_{z}=0.9t$, the energy gap
($E_{\textrm{gap}}=2E_{1}$) is always nonzero and increases monotonically
with disorder strength, as anticipated for a disordered conventional
$s$-wave superfluid \cite{Ghosal2001}. In contrast, at a large Zeeman
field $h_{z}=1.8t$, there are two zero-energy modes that correspond
to the Majorana fermions localized at the two edges. By increasing
the disorder strength above a threshold $V_{c}^{*}$, the zero-energy
modes crease to exist \cite{Cai2012,Wang2015}, suggesting the loss
of the non-trivial topological property of the superfluid. The two
critical disorder strengths, $V_{c}$ and $V_{c}^{*}$, do not necessarily
take the same value. 

\begin{figure}
\begin{centering}
\includegraphics[width=0.4\textwidth]{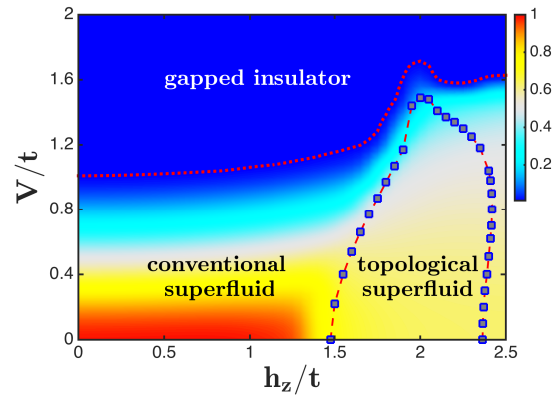} 
\par\end{centering}

\protect\caption{(color online) Phase diagram of a 1D spin-orbit coupled Fermi gas
in bichromatic optical lattices at $U=4.5t$ and $\lambda=2t$. The
color bar indicates the superfluid stiffness. A vanishingly small
superfluid stiffness (i.e., the red dotted line) determines the phase
transition to a normal state ($V_{c}$), due to the many-body Anderson
localization of the ground state. The blue squares with dashed line
($V_{c}^{*}$) enclose the phase space, where the superfluid is topologically
non-trivial and Majorana fermions exist at the two open edges. The
other parameters are the same as in Fig. \ref{fig3}.}

\label{fig4}
\end{figure}
By collecting the two critical disorder strengths at different Zeeman
fields, we arrive at our main result - the phase diagram of a spin-orbit
coupled Fermi gas in quasi-random lattices - as illustrated in Fig.
\ref{fig4}. It is remarkable that the many-body localization strength
$V_{c}$ (i.e., the red dotted line) follows very closely behind the
two-particle localization strength $V_{c,2p}$. This indicates that
the loss of coherence between Cooper pairs and the localization of
pairs occurs nearly at the same time. It is also interesting that,
in the presence of disorder, we find a large parameter space to accommodate
topological superfluids. In particular, a topological superfluid appears
to be more robust against disorder than a conventional $s$-wave superfluid.
There is no direct transition from a topological superfluid to the
gapped insulator. Therefore, the scenario (b) in the schematical phase
diagram Fig. 1 is favored.

\begin{figure}
\begin{centering}
\includegraphics[width=0.45\textwidth]{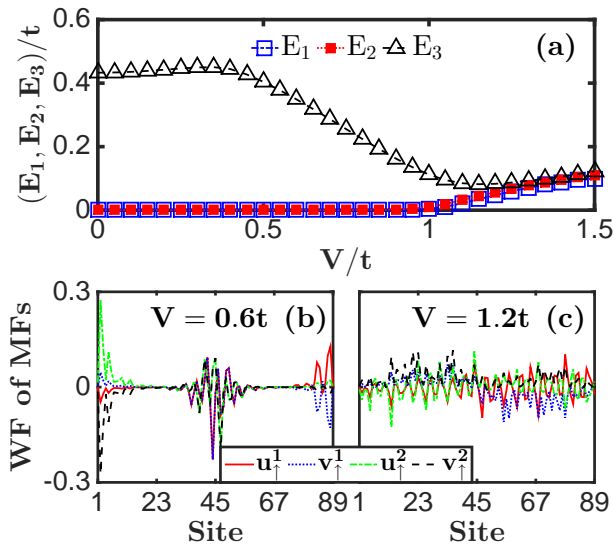} 
\par\end{centering}

\protect\caption{(color online) Upper panel: Evolution of the three lowest quasiparticle
excitation energies as a function of the disorder strength, under
an open boundary condition and in the presence of a soliton at the
center of the system. Lower panel: The wave functions of the two Majorana
fermions ($E_{1}$ and $E_{2}$) at the disorder strength $V=0.6t$
(b) and $V=1.2t$ (c). Here, we fix the Zeeman field strength to $h_{z}=1.8t$.
The other parameters are the same as in Fig. \ref{fig3}.}

\label{fig5}
\end{figure}

\textit{Anderson localization of Majorana fermions}. ---We now examine
in more detail the localization of Majorana fermions, which gives
the critical strength $V_{c}^{*}$. Experimentally, it seems convenient
to create and manipulate Majorana fermions by engineering solitons
via the phase-imprinting technique \cite{Yefsah2013,Ku2015}. At the
site of each soliton, there exist two soliton-induced Majorana fermions
\cite{Xu2014,Liu2015}. In Fig. \ref{fig5}, we plot the energy and
wavefunctions of low-lying quasiparticle modes in the presence of
a dark soliton at the center of the topological superfluid and under
an open boundary condition \cite{footnote2}. The wave-functions of
Majorana fermions are essentially not affected by a weak or moderately
strong disorder (Fig. \ref{fig5}b) and are well localized at the
edges or at the site of soliton. Towards the critical strength $V_{c}^{*}$
(Fig. \ref{fig5}c), however, the wave-functions spread over the whole
lattice sites. Majorana fermions disappear, although the quasiparticle
states are still extensive. 

\textit{Conclusions}. ---We have proposed to study the Anderson localization
of a spin-orbit coupled Fermi gas in one-dimensional quasi-disordered
lattices. In the absence of disorder, the system features a conventional
$s$-wave-like superfluid and topological $p$-wave-like superfluid
at small and large Zeeman fields \cite{Zhang2008}, respectively.
By tuning the Zeeman field, we have investigated how these superfluid
states lose their superfluidity with increasing disorder strength.
We have found that topological superfluids (and hence the hosted Majorana
fermions) are very robust against disorder and they lose their non-trivial
topological properties before finally turn into gapped insulators.
We have complemented our many-body study by considering the localization
of a single Copper pair. The result indicates that the localization
of pairs and the loss of coherence between pairs occurs simultaneously. 

In the near future, it is of interest to consider Anderson localization
of the spin-orbit coupled Fermi gas system in two-dimensional disordered
lattices \cite{Bordia2015}. The strong quantum phase fluctuations
near topological and localization transitions could be addressed by
using a zero-temperature Gaussian fluctuation theory \cite{Hu2006,Diener2008,He2015}.
\begin{acknowledgments}
GX was supported by the NSF of China (Grant Nos. 11374266 and 11174253)
and the Program for New Century Excellent Talents in University. XJL
and HH were supported by the ARC Discovery Projects (Grant Nos. FT130100815,
DP140103231, FT140100003, and DP140100637) and NFRP-China (Grant No.
2011CB921502). \end{acknowledgments}


\begin{thebibliography}{10}
\bibitem{footnote2}Theoretically, a dark soliton at a site $A$ can
be created by imposing the following constraint to the local pairing
gap: $\Delta_{i<A}>0$, $\Delta_{i=A}=0$, and $\Delta_{i>A}<0$.
\end{thebibliography}
\end{document}